*IEEE* Access

## RESEARCH ARTICLE

# SAppKG: Mobile App Recommendation Using Knowledge Graph and Side Information-A Secure Framework

DAKSH DAVE[1], ADITYA SHARMA[2], SHAFI'I MUHAMMAD ABDULHAMID[3], ADEEL AHMED [4], ADNAN AKHUNZADA [5], (Senior Member, IEEE), AND RASHID AMIN[6]
[1]Department of Electrical Electronics, Birla Institute of Technology And Science, Pilani, Rajasthan 560103, India
[2]Department of Computer Science and Information Science, Birla Institute of Technology And Science, Pilani 560103, India
[3]Department of Information Technology, Science and Technology Division, Community College of Qatar, Qatar
[4]Department of Information Technology, The University of Haripur, Haripur, Khyber Pakhtunkhwa 22620, Pakistan
[5]Department of Computer Sciences, Technical University of Denmark, 2800 Kongens Lyngby, Denmark
[6]Department of Computer Sciences, University of Chakwal, Rawalpindi 48800, Pakistan

Corresponding authors: Shafi'i Muhammad Abdulhamid (shafii.abdulhamid@ccq.edu.qa) and Daksh Dave (f20180391p@alumni.bits-pilani.ac.in)

This work was supported by the Qatar National Library.

**ABSTRACT** Due to the rapid development of technology and the widespread usage of smartphones, the number of mobile applications is exponentially growing. Finding a suitable collection of apps that aligns with users' needs and preferences can be challenging. However, mobile app recommender systems have emerged as a helpful tool in simplifying this process. But there is a drawback to employing app recommender systems. These systems need access to user data, which is a serious security violation. While users seek accurate opinions, they do not want to compromise their privacy in the process. We address this issue by developing SAppKG, an end-to-end user privacy-preserving knowledge graph architecture for mobile app recommendation based on knowledge graph models such as SAppKG-S and SAppKG-D, that utilized the interaction data and side information of app attributes. We tested the proposed model on real-world data from the Google Play app store, using precision, recall, mean absolute precision, and mean reciprocal rank. We found that the proposed model improved results on all four metrics. We also compared the proposed model to baseline models and found that it outperformed them on all four metrics.

## I. INTRODUCTION

With the development of hardware technologies, a significant increase has also been observed in the number of mobile applications available to the user to perform a particular task. Due to the availability of such a large number of apps, it becomes a challenge for a user to select an appropriate set of apps that satisfies the user's needs. Recommender systems are based on information filtering and are traditionally classified into content-based filtering (CBF), collaborative filtering (CF), and hybrid methods [1]. In the content-based filtering (CBF) approach, items chosen for recommendation

The associate editor coordinating the review of this manuscript and approving it for publication was Wei Liu.

are based on the user's current preferences and the similarity of the items. However, CBF suffers from the problem of overspecialization [2], which means that it cannot diversify the existing interests of users because it can only provide recommendations based on the current interests of the user. Collaborative Filtering(CF), on the other hand, takes into account a user's prior interactions and searches for similar users to make recommendations based on common preferences. But the downside of CF is that it is difficult to add any item's side features to increase model quality. Side features, also known as side information, are any additional information about the input data, such as user/item features, images, text, etc., that, if included, can significantly improve the model's performance. For example, in mobile app suggestions, side information of an app may include app category, age limit, installations, user ratings, and so on.

Around 3.29 million apps in the Google play store and over 2.1 million apps in the Apple app store are available as of May 4, 2022 [3]. A simple search for "calculator" in the Google Play store yields more than 100 results for the user. Rather, a person installs, on average, 35 apps out of these millions of apps available in app stores [4]. The result is a high sparsity in user-app interaction data. Side information compensates for the sparsity by providing additional data and features about user-app interactions. Researchers have tried to include as much side information as possible in the app recommendation algorithms; however, they have only been able to use a limited number of side information types. Furthermore, researchers treated the side information as separate user and app properties, ignoring the relationships and semantics between them [5].

Guo et al. [6] used only the apps' category, and Cao et al. [7] only took into account the version of an app. Therefore, properly integrating and utilizing side information to produce good app recommendations is a topic in exploration that faces difficulty in the current research world. The ultimate goal of an app recommender system is to identify a small set of apps that meet the user's personal interests and requirements. To accomplish this, app recommender systems must gather user information in one of two ways: users voluntarily providing app preferences and requirements, or from users' mobile devices using installed app collection, sensors, phone records, contacts, emails, web browsing history, and several other [8]. While it helps app recommender systems provide more personalized recommendations, it also compromises users' privacy. Recommending apps to users without directly infringing on their privacy is a difficult task.

Existing approaches [1], [6], [9], [10], [11], [12], [13] were unable to account for the multi-relational structure of app recommendations. Knowledge graphs are better suited for such scenarios since knowledge graphs have a heterogeneous structure that allows for the inclusion of such complex interactions between items and incorporating side information. The fundamental benefit of knowledge graph-based models is their ability to represent diverse, complicated, unstructured data using rich ontologies and semantics, making the knowledge graph a commonly used tool in recommendation systems [14]. Zhang et al. [5] created a knowledge graph to deliver app suggestions while adding side information such as content topic, app size, and app popularity. However, a fundamental disadvantage of this approach is that it makes use of user data from the user-app interaction matrix, which raises privacy concerns.

To resolve the aforementioned challenges, we present SAppKG, a knowledge graph-based secure method- an ontology for making app recommendations without specifically utilizing any user information. We developed a knowledge graph using only side information from apps, such as content rating, app genreID, etc. More specifically, we proposed two frameworks, SAppKG-S and SAppKG-D, for securely and privately recommending new apps using a knowledge graph.

In this paper, our contributions are:
- We propose a mobile app recommendation framework called SAppKG based on similar apps without accessing the user's data to ensure user privacy and solve the problem of data sparsity by incorporating side information.
- We proposed SAppKG-S, a shallow embedding-based secure model for app recommendation that employs a range of shallow embedding models such as TransD, TransH, and Complex.
- We also presented a novel model, SAppKG-D, which integrates shallow and deep embedding techniques and employs TransD to identify unique embeddings for individual nodes. These embeddings are then combined with those produced by a graph convolution network to generate recommendations for apps based on the acquired embeddings.
- We also introduced a support metric that measures the relatedness between two relations.
- We collected real data from the Google Play store for capturing the relationships and side information about app connections.
- We compared a proposed framework with baselines and evaluated the proposed framework on the basis of precision, recall, mean absolute precision, and "Mean Reciprocal Ranking and found improved results.

The rest of the paper is organized as follows: Section II contains the literature review, Section III discusses the methodology, Section IV discusses the results, and Section V lists references.

## II. RELATED WORK

We divided our literature review based on mobile app recommendation and security in mobile app recommendation.

### A. MOBILE-APP RECOMMENDATION

Given the importance of mobile app recommendation for both users and platforms, several mobile app recommendation approaches have been developed. Committee [9] created AppJoy, a personalized app recommendation system that proposes mobile apps based on the user's app usage history (by examining users' current app consumption habits), a client-server architecture, and a collaborative filtering algorithm. Liang et.al. [10] focused on exploiting the permissions and functionalities needed for an app, arguing that the combination of permissions, functionality, and user interests of the app plays an important role in recommending a personalized app. The authors offer the App Risk Score Method(ARSM), which reflects the app's trustworthiness based on a relation between app permissions and user ratings. Then, a modified matrix factorization algorithm, MFPF (Matrix Factorization Algorithm Based on Permissions and Functionalities), predicts a rating for a new app from a specific user based on the user's interest in that app and permission similarities with another app. Tu et al. [13] present IMCF+(Interest-aware matrix co-factorization plus), a collaborative filtering approach based on the assumption that when a user posts on social media about what they like, they are more likely to install related apps on their mobile devices. IMCF+ is a transfer learning-based technique that uses user-app usage data, user tweet/post data, and app-to-tweet word correlation data and generates a personalized ranking for unseen apps per user. Whereas, Guo et al. [6] claims that collaborative filtering (CF) and matrix factorization (MF) techniques suffer from poor feature extraction, prohibiting them from correctly utilizing the acquired features. Authors offer KDFM(Knowledge-based Deep Factorization Machine), which uses categorical (app name, user name, etc.) and textual knowledge(user reviews, app description, etc.) to estimate user ratings for mobile apps. The distinctive feature of this technique is the implementation of an attention-aware deep encoder to map textual knowledge into the topic-based dense representation. Liang et al. [15] advance the notion of attention-based app recommendation by proposing MV-AFM (Multi-view Attentional Factorization Machines), which analyses the associations between features from various views(represented by a set of features) using the attention mechanism. This method introduces two-tier attention networks and separates feature interactions based on various views. A key aspect of MV-AFM is the ability of two attention sub-networks to distinguish between feature weights inside each view and interactions across views. Xu et al. [12] claim that two users would have similar app preferences if their contextual app usage patterns were the same. Authors feed a user-app interaction matrix into a neural network with two components: an app context prediction module and a user preference prediction module. The former provides context for an app, whereas the latter forecast a user's app preferences. Contextual data from app usage patterns enhance the prediction of user preferences over applications, and user preferences for each app provide the recommended apps. Maherswari et.al. [16] signifies the importance of recommending a proper version of the app to the user. The authors propose integrating Probabilistic Matrix Factorization (PMF), an advanced and more powerful matrix factorization approach, with the Version Evolution Progress Model(VEPM). PMF is used to find the latent feature representation for the user and app, whereas VEPM represents the evolution of mobile app versions and trains a model to improve rating prediction performance.

KGEP (Knowledge Graph Convolutional Embedding Propagation Model) [5] is a major technique that uses knowledge graphs to provide app recommendations to users. The innovation of this technique is that it incorporates side information1 about relationships into a knowledge graph (KG) based on the link between users and apps. Furthermore, final vectors for users and apps are generated using a technique similar to KGCN2 embeddings [17] and merged with generic KG embeddings(TransD embeddings). The probability value of a user acquiring a certain app is then computed using these vectors to provide final app recommendations. A recent study on app recommender systems by Tejaswi. et. al. [18] proposed a unique Multi-Criteria Mobile App Recommender System (MCMARS) model that recommends the top apps to the users as well as helps the developers in improving their app performance through intelligent recommendations.

### B. SECURITY IN MOBILE APP RECOMMENDATION

To provide more personalized recommendations, app recommendation systems often access user information in some way. It seriously breaches user privacy. In their study, Sandhu et al. [8] highlights three important aspects of mobile recommendation systems: the sources used to acquire user data, the privacy issues raised by various data collection techniques, and remedies to these problems. The authors also discuss the privacy-personalization trade-off, which forces users to decide between privacy and personalized suggestions. Highly personalized recommendations mean privacy must be compromised, but less accurate recommendations arise from putting more importance on privacy. Additionally, there are instances when users must decide whether to accept the privacy policy established by mobile recommendation systems or reject it and avoid using the system altogether. Furthermore, because the privacy policies of mobile recommender systems are written in a technical and complex manner, it is challenging for the typical user to understand them, and users are uninformed of the magnitude of privacy loss. Beg et al. [19] give a detailed overview of the numerous methods suggested to ensure privacy and security in the recommendation of mobile apps. According to the authors, the availability of multiple sensors and the permissions granted to third-party programs on mobile devices make privacy problems in mobile devices more detrimental. The authors mention methods implemented to protect privacy: correlation-based, deep neural network adversarial learning, encryption-based, perturbation and noise-based, differential privacy-based, and homographic encryption. Additionally, Ravi et al. [20] suggest the implementation of a secure framework, SECRECSY, to facilitate recommendation systems that have migrated their data and infrastructure to a cloud-based platform. This measure upholds users' confidentiality throughout the process of generating recommendations.

Zhu et al. [21] suggested a recommendation system for apps that can analyze the level of privacy of an app depending on the permissions it seeks. The authors then recommend apps while trying to balance an app's popularity and users' apprehensions about security. This approach has the drawback of being non-personalized and is unable to offer customized app suggestions based on users' needs and interests. Xu et al. [22]propose PPMARS-C (privacy-preserving mobile app recommendation system for cloud services) and PPMARS-S (privacy-preserving mobile app recommendation system for social networks) as two mobile app recommendation models that preserve privacy. While the latter is a distributed system that makes suggestions in social networks, the former is a centralized system that delivers app recommendations for a cloud service environment. Both approaches employ data regarding users' trust behaviours with regard to mobile apps they have downloaded and installed. To ensure the security of identification, data transfer, and data processing, public-key encryption and homomorphic encryption are utilized.

Recent studies [23], [24] have tried to address the privacy challenges in the recommendation systems and developed a qualitative approach through extensive peer-review of articles for helping the researchers adopt the best approaches in the recommenders systems for resilience in privacy and security for risk mitigation.

Compared to the approaches mentioned above, the advantages of our method are twofold. The first thing to note is that each of the above methods makes use of data on how the user and the app interacted. Second, despite the ability of knowledge graphs to deliver rich information relating to data, little effort has been made in the area of mobile app recommendations using knowledge graphs.

## III. METHODOLOGY

Our framework, depicted in Figure-3, comprises three modules that collectively enable effective app recommendation. The first module, Data Scraping, and Entity Identification, involves extracting app data and side information from JSON files, followed by entity identification through preprocessing. The second module, Constructing a Knowledge Graph and KG Quality Check, employs four techniques to generate a knowledge graph (KG) representing app data, and capturing entity relationships. A thorough quality check assesses the statistical characteristics and relation scores of the KG. The third module encompasses three subsections: Initializing the KG Embeddings utilizes the SAppKG-S model to initialize embeddings encoding semantic information and relationships. The subsequent subsections, SAppKG-S and SAppKG-D, focus on training and evaluating KG-based recommendation models using shallow and deep learning techniques. The KG embeddings play a vital role in enhancing recommendation accuracy and performance.

### A. DATA SCRAPING AND ENTITY IDENTIFICATION

We created a dataset of real mobile apps from the Google Play store that includes information about the apps from three major categories—*Photography*, *Productivity* and *Games*—and extracted data for 200 apps from each category. The extracted data contained information about thirteen app attributes. These are-*appId*, *adSupported*, *contentRating*, *editorsChoice*, *genreId*, *installs*, *offersIAP*, *ratings*, *released*, *reviews*, *scoreText*, *size*, and *video*. Then, in order to represent extracted data in structural form and capture rich semantic information, we build a knowledge graph (KG).

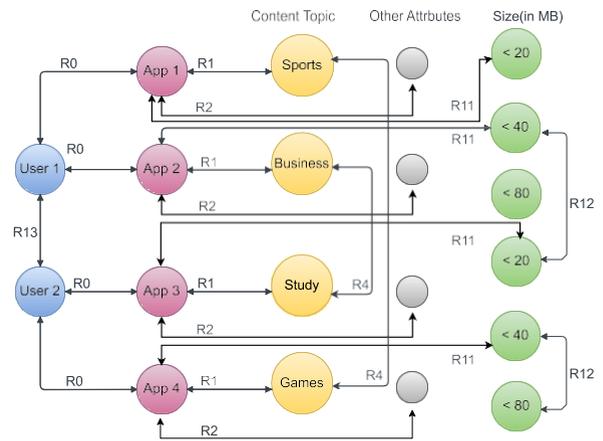

**FIGURE 1.** User-App KG.

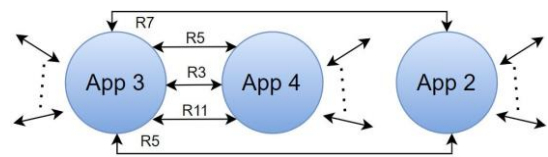

**FIGURE 2.** SAppKG: Knowledge graph.

To build the KG we pre-processed the data of thirteen selected attributes using techniques such as-

- Normalization: This included eliminating redundant and unstructured data and making the data appear similar across all records and fields. Example: Size
- Quantile mapping: Quantile mapping defines the bins using percentiles based on the distribution of the data, not the actual numeric edges of the bins. The data is divided into a set of quantiles and we have mapped the data falling within a similar quantile range to a single value. Examples: Ratings, Reviews, ScoreText
- Interval mapping: It maps the data to a bin if the data falls within the bin interval range.Examples: Installs, Released, Size
- Category mapping:It puts the data having a similar set of categories into the same bin. Examples: AdSupported, Editor's Choice, offersIAP, video, ContentRating, GenreId

The appId is considered as the nodeId, and the remaining twelve properties are chosen as node attributes. Based on the twelve different node attributes mentioned above, we constructed twelve relations between the nodes (i.e., apps). If the node attribute values of the two apps are similar, a certain relation connects the two apps. For example, if the *ContentRating* of the two apps $a_i$ and $a_j$ was similar, the relation CRSIMILAR was established between them. Table 1 summarizes the twelve relations that were created based on various attributes of the nodes and the corresponding side information that each node possesses.

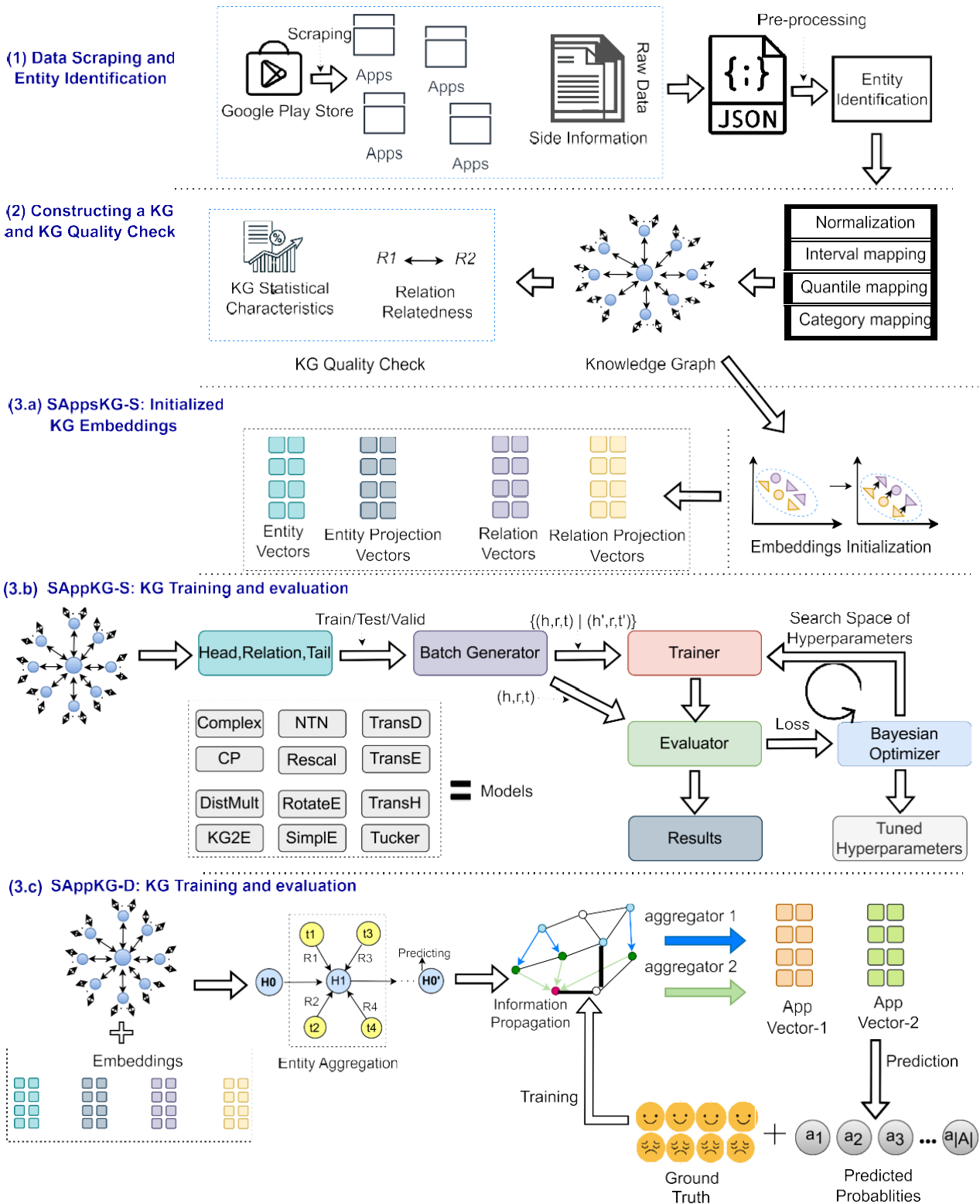

**FIGURE 3.** Proposed framework for SAppKG.

## B. CONSTRUCTING A KNOWLEDGE GRAPH

### 1) PRIVACY AND SECURITY

To protect user privacy and ensure security, our model adopts a privacy-preserving strategy that avoids direct usage of the user dataset. Mobile app attributes can be broadly categorized into two types: design-specific attributes and user-specific attributes. Design-specific attributes refer to characteristics that are available at the time of app launch and remain constant, such as the app's size, ad support, pricing model (free or paid), and compatibility requirements. These attributes provide insights into the app's structure, features, and monetization strategy. On the other hand, user-specific attributes capture information that evolves over time based on user interactions with the app. These attributes include the number of installs, user reviews, ratings, app score, and other user-generated feedback. User-specific attributes reflect the app's popularity, user satisfaction, and overall performance.

**TABLE 1.** Head relation tail table.

| Relation No. | Relation | Head Feature | Tail Feature | No. of Groups | Related Side Information |
|---|---|---|---|---|---|
| 0 | ADSIMILAR | AD | AD | 2 | App supports advertisements |
| 1 | CRSIMILAR | CR | CR | 4 | App's content rating data |
| 2 | ECSIMILAR | EC | EC | 2 | App has editor's choice |
| 3 | GIDSIMILAR | GId | GId | 3 | App's genre Id data |
| 4 | INSSIMILAR | Installs | Installs | 4 | Number of app installs data |
| 5 | IAPSIMILAR | IAP | IAP | 2 | App supports In-App-Purchases |
| 6 | RTGSIMILAR | Ratings | Ratings | 5 | Number of users who have rated the app |
| 7 | RELSIMILAR | Released | Released | 7 | App's released date |
| 8 | REVSIMILAR | Reviews | Reviews | 5 | Number of users who rated as well as written reviews |
| 9 | STSIMILAR | ScoreText | ScoreText | 8 | Mean rating of all the users who rated the app |
| 10 | SSIMILAR | Size | Size | 5 | App's size entity data |
| 11 | VSIMILAR | Video | Video | 2 | App supports video |

Fig. 1 illustrates the structure of the knowledge graph proposed by Zhang. et. al. [5], which encompasses user nodes, app nodes, and app attribute nodes. These nodes are interconnected through various relations, representing the associations between them. In our proposed methodology, we prioritize the use of app nodes as the primary entities in the construction of the knowledge graph. To achieve this, we employ transformation and encoding techniques to effectively incorporate the side information associated with apps within the graph's relations. By adopting this approach, we are able to address important concerns such as user privacy preservation and prevention of information leaks, especially in the event of a knowledge graph leak. Furthermore, this methodology offers the additional benefit of reducing the overall number of nodes in the knowledge graph, optimizing its structure, and enhancing its efficiency.

The process of our KG construction is discussed in detail in the next section. In our graph, we leverage a combination of both design-specific(size, Ad Support, free or paid, etc) and user-specific(installs, reviews, ratings, app Score, etc) interaction attributes available in the dataset. Fig. 2 gives a glimpse of our Knowledge Graph that gets built on completing the KG construction processes. We can see that the same apps can be connected to a single app by multiple relations and using the same relation a single app can be connected to multiple apps. We group the values into appropriate categories and interconnect the nodes based on their category similarities to build the KG.

To evaluate the effectiveness of our model in app-app recommendations, even without relying on the user dataset, we conduct a comparative analysis with the KGEP model [5] as shown in Table 12. Through this performance evaluation, we quantitatively measure the efficiency and efficacy of our proposed SappKG-D model, showcasing its ability to deliver accurate recommendations. The experimental results section presents a detailed and in-depth analysis of the table, providing insights and observations derived from the comparison.

### 2) BUILDING A KG
The subsequent step involves building the relationships, so we needed to model these relationships into a tangible graph.

**TABLE 2.** Relation (0,2,5,11) - Building(AdSupported, editorsChoice, offersIAP, video).

| Attributes | adSupported | editorsChoice | offersIAP | video |
|---|---|---|---|---|
| TRUE | 956 | 44 | 926 | 726 |
| FALSE | 837 | 1749 | 867 | 1067 |

All the relationship attributes mentioned in Table 2 above have a binary value mapping True and False. We use Category mapping and group the True values into one category and the False values into another category. We interconnect the *appId's* with similar category value with relation-0.

**TABLE 3.** Relation 1-Building (Content Rating and GenreId).

| ContentRating | | GenreId | |
|---|---|---|---|
| Everyone | 1369 | Photography | 556 |
| Teen | 267 | Productivity | 482 |
| Everyone 10+ | 113 | Games(Total) | 755 |
| Mature 17+ | 44 | | |

The Table 3 shows the relationship attributes for *contentRating* which has four values- *Everyone, Teen, Everyone 10+* and *Mature 17 +.* We use Category mapping and group the *Everyone* values into one category and repeat the same for the other values. The genre has three main values-*Photography, Productivity,* and *Game* as shown in the Table 3. All the sub-categories of the *GAME* were grouped into a single category, *GAMES,* to reduce the feature space. We use Category mapping on the obtained labels

Installs attribute has 18 values- With the help of plotting we categorize these 18 groups into four groups- *(0 - 500+), (1,000+ - 5,0000+), (1,00,000+ - 50,00,000+),*

*(1,00,00,000+ - 500,00,00,000+)*. We use Interval mapping to group the values present in a particular interval into its interval-specific categories. Fig. 4 shows the install key-value pair plot.

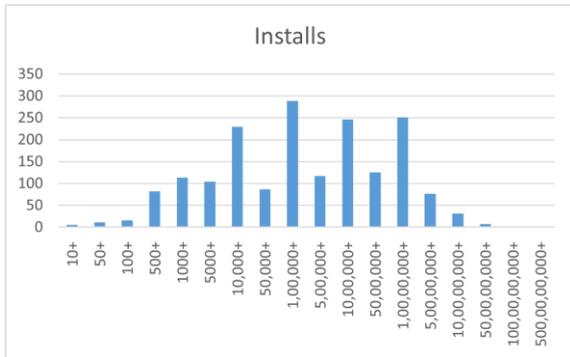

**FIGURE 4.** Installs key-value plot.

**TABLE 4.** Relation 6,7,8,9,10-Building (Ratings, Released,ScoreText,Size).

| Attributes | Ratings | Released | Reviews | ScoreText | Size |
|---|---|---|---|---|---|
| count | 1789 | 1784 | 1789 | 1789 | 1793 |
| mean | 561222 | 1267 | 25005 | 4.053158 | 55352 |
| std | 3587513 | 1135 | 1742890 | 0.883409 | 90369 |
| min | 0 | 17 | 0 | 0 | 0 |
| 25% | 277 | 208 | 127 | 3.9 | 7680 |
| 50% | 8228 | 971 | 3154 | 4.3 | 29696 |

Table 4 has four relations. The relation *Ratings* and *Reviews* has a continuous distribution of values. We use *qcut* as a Quantile-based discretization function that divides up the underlying data into equal-sized bins. The function defines the bins using percentiles based on the distribution of the data, not the actual numeric edges of the bins. So, there are a total of five quantile ranges *0-0.2,0.2-0.4,0.4-0.6,0.6-0.8,0.8-1.0*. The quantiles group the values into the appropriate quantile by assigning the value in between the quantile range and accordingly labelling it between *0 - 4*.

The relation *Released* has a continuous distribution of dates. We subtract the Released date from today's date to get the number of days *(Number of days = Today's date - Released)*. We use interval mapping and put these values into intervals, grouping them as released within *1 month, 2 month, 3 month, 4 month, 5 month, 6 month, 7 month, 8 month, 9 month, 10 month, 11 month, 12 months* and *released after a year*. The relation *ScoreText* has a discreet set of values grouped from *0.0* to *5.0* constituting fifty intervals. We use *qcut* to divide the data into 8 quantile ranges to differentiate them The relation Size has a continuous distribution of values in an object format. We normalize the data into a *Kb* format. We then use interval mapping and put these values into intervals, binning them into values *0-6*. The bins are labelled as *(0-1)-0, (1-20000)-1, (20000-40000)-2, (40000-60000)-3, (60000-80000)-4, (80000-100000)-5, (100000-2000000)-6*. The apps that take less than *1 kb* space, *vary with the device*,

or have no size specifications have been grouped into a single category. The next category is of apps with sizes less than *20 Mb, 40Mb, 60Mb, 80Mb, 100Mb*. The last category is apps with more than *100 Mb*.

### 3) KNOWLEDGE GRAPH QUALITY CHECK
In order to examine the issues encountered while designing a Knowledge Graph we examine its statistical characteristics helping us get a close idea about its structural properties [25].

To measure the interdependence between the different relationships, Schanbel et al. used a cosine-based similarity of the embeddings for two words to measure the human relatedness scores and present a novel evaluation framework based on direct comparisons between embeddings. They developed a novel Coherence task to measure the intuition that neighbourhoods in the embedding space should be semantically or syntactically related [26]. To measure the relatedness, we measure the relatedness score between every two relations. To compute the relatedness between labels $r_i$ and $r_j$, we first define the support of $r_i$, $r_j$ as, as shown in (1):

$$supp(r_i \rightarrow r_j) = \frac{|r_i \cap r_j|}{|r_i|} \qquad (1)$$

$r_i$ and $r_j$ denote the number of nodes that have a relation $r_i$ or $r_j$ and $r_i \cap r_j$ denotes the number of nodes that have both relations $r_i$ and $r_j$. The support function is not symmetric, inspired by the definition of F-measure, the relatedness of $r_i$ and $r_j$ is defined as follows in (2):

$$R(r_i \rightleftarrows r_j) \frac{(2 \times supp(r_j \rightarrow r_j) \times supp(r_i \rightarrow r_j))}{(supp(r_i \rightarrow r_j) + supp(r_j \rightarrow r_i))} \qquad (2)$$

The learning process is augmented with the relational similarity relatedness matrix giving a brief view of the similar relation pairs. The Table 5 given below lists out the different relatedness scores in between the different relations in the form of a table where *relation-i* to *relation-j* boxes contain their relatedness scores in between the two relations $i$ and $j$.

The given Table 5 shows the relatedness scores between different relations based on the support function defined earlier. The relatedness scores range from 0.9949 to 1.0, where a score of 1.0 indicates the highest relatedness between two relations. We can see that most of the relations are highly related to each other in the graph, with a score of 1.0 occurring twenty-four times. The relations with the least relatedness scores are 1-7 and 4-7, both having a score of 0.9949 and 0.9957, respectively, with a frequency of one. These scores provide valuable insights into the interdependence between different relations and can be used to optimize the knowledge graph for better performance.

Table 5 describes the 11 different statistical attributes that we calculate and examine to view the Graph properties.

### C. EMBEDDINGS BUILDING AND TRAINING A KNOWLEDGE GRAPH
Our Knowledge Graph contains a set of triples in the form of *(h,r,t)* represented as Facts. Fig. 3 lists out the flow of our study. The proposed framework consists of four modules as shown in Fig. 3; the first module is related to Data Collection and Preprocessing, the second module is Building a Knowledge Graph, the third module is Building Node Embeddings, the fourth is using shallow embedding methods to build and train the KG on different embeddings using the SAppKG-S model, and the fifth is using SAppKG-D; A deep embedding based approach for embeddings propagation through a modified KGCN.

**TABLE 5.** KG statistical characteristics.

| Relatedness Score | Relations |
|---|---|
| 0.9949 | 1-7 |
| 0.9957 | 4-7 |
| 0.9968 | 7-8, 7-9, 6-7 |
| 0.9971 | 4-6, 4-8, 4-9 |
| 0.9974 | 0-7, 2-7, 3-7, 5-7, 7-10, 7-11 |
| 0.9982 | 0-4, 2-4, 1-4, 3-4, 4-5, 4-10, 4-11 |
| 0.9988 | 0-6, 0-8, 0-9, 1-6, 1-8, 1-9, 2-6, 2-8, 2-9, 3-6, 3-8, 3-9 |
| 0.9988 | 6-10, 6-11, 5-6, 5-9, 8-10, 8-11, 9-10, 9-11 |
| 1 | 0-1, 0-2, 0-3, 0-5, 0-10, 0-11, 2-3, 2-5, 2-10, 2-11, 1-2, 1-3 |
| 1 | 1-5, 1-10, 1-11, 3-5, 3-10, 3-11, 6-8, 6-9, 5-10, 5-11, 8-9, 10-11 |

| Parameter | Values | Description |
|---|---|---|
| Nodes | 1793 | The number of nodes in the Graph |
| Edges | 21433 | The number of edges in the Graph |
| Density | 0.00667 | The Density or sparsity of the graph |
| Average Degree | 11.95 | Degree of the Graph |
| Multiplex Dyads | 1793 | Number of nodes that are connected by multiple relations |
| Triads Possible | 959097216 | Number of Triangles found in the graph |
| Otriads | 21421 | Number of Open Triangles found in the graph |
| Degree Variance | 35.8 | A variance in the degree of the nodes in the graph |
| Edge Disconnecivity | 11 | Min edges to remove to make disconnected graph |
| APPFDIAM | 22 | Maximum Shortest Path between every pair of vertices |
| Average Shortest Path Length | 11.215 | Shortest path between every pair of vertices |

The two proposed methods SAppKG-S and SAppKG-D are described below.

### 1) SAppKG-S

To generate node embeddings, most approaches rely on a shallow embedding method that performs an embedding lookup to map nodes to their embeddings. Each node is then trained to produce a unique embedding. In this paper, we introduce the SAppKG-S method, which also employs a shallow embedding approach for recommending apps in a knowledge graph. We compare the performance of various shallow embedding models in the SAppKG-S framework.

We use Pykg2vec, a library built on PyTorch that learns the entities and relations representation. State-of-the-Art Knowledge Graph Embedding algorithms have been implemented using this library. Table 7 shows the different embeddings models used in our study categorized on the basis of their model training into 3 categories:

1) Pairwise (margin) based Training KGE Models

- NTN [27]: The Neural Tensor Network (NTN) is a neural embedding model designed for Knowledge Graphs. It incorporates second-order correlations into nonlinear neural networks. The score function

is given in (3):

$$f_r(h, t) = u_r^T f(h^T) \widehat{W}_t + M_r \begin{bmatrix} h \\ t \end{bmatrix} + b_r \quad (3)$$

where $\widehat{W}_r$ represents 3-way tensor, $M_r$ is a weight matrix, $b_r$ is bias and $f()$ is $tanh$ function.

- TransE [28]: It computes the similarity of a pair of entities and the relationship between them using a basic distance-based approach. If the triplet of head, relationship, and tail (h, r, t) holds, then the aggregate of the embeddings of h and r should be as close as possible to the embeddings of t. It follows the scoring function as in (4):

$$f_r(h, t) = ||h + r - t||_2^2 \quad (4)$$

- TransH [29]: The TransE embeddings exhibit limitations in their ability to effectively handle multi-relational graphs. The aforementioned constraint was successfully addressed through the implementation of the TransH model. The TransE model incorporates hyperplanes as a means of projecting the embeddings of entities h and t onto relation-specific hyperplanes. The scoring function followed by TransH is given in (5):

$$f_r(h, t) = ||(e_h - w^T e_h w_r) + d_r - (e_t - w^T e_t w_r)||_2^2 \quad (5)$$

where $h_\perp$ and $t_\perp$ are projections of head and tail on the hyperplane of r.

- TransD [30]: The TransD model employs a dual-vector representation for both entities and

---

**Algorithm 1** Knowledge Graph Embedding and Training With SAppKG-S

---

**Require:** Knowledge Graph $G = (E, R)$, train/validation/test split $G_{train}$, $G_{valid}$, $G_{test}$, KGE algorithms $\mathcal{A}$, hyperparameters $\vartheta$
**Ensure:** Trained KGE model and evaluation metrics on $G_{test}$
  1: Generate embeddings for each entity and relation in $G$ using pykg2vec
  2: Train each KGE algorithm $A \in \mathcal{A}$ on $G_{train}$ using the generated embeddings
  3: Select the best performing KGE model using $G_{valid}$
  4: Generate embeddings for $G$ using the best KGE model and hyperparameters
  5: Use the embeddings to predict missing links in $G_{test}$ and calculate evaluation metrics
  6: **procedure** Embeddings Generation
  7:     Create instance of KnowledgeGraph and call `prepareData()` to pre-process data
  8:     Import desired KGE algorithm and its configuration using `Importer` class
  9:     Create instance of the algorithm and call `trainModel()` to generate embeddings
 10: **end procedure**
 11: **procedure** Training
 12:     Tune hyperparameters $\vartheta$ for the chosen KGE algorithm using Bayesian optimization
 13:     Create an instance of the algorithm with the best $\vartheta$
 14:     Train the algorithm on $G_{train}$ using the embeddings and `Trainer` class
 15: **end procedure**
 16: **procedure** Evaluation
 17:     Generate embeddings for $G_{test}$ using the trained algorithm
 18:     Use the embeddings to predict missing links in $G_{test}$
 19:     Calculate evaluation metrics (e.g. MR, hits@k, MRR) for predicted links and compare to ground truth
 20: **end procedure**

---

relations. One vector to describe the entity's meaning and another to dynamically build a mapping matrix $M_{rh}$, $M_{rt}$. These matrices project head and tail entities from entity space to relation space. The score function is shown in (6):

$$f_r(h, t) = -||(r_p e^T + I)e_h + r - (r_p e^T + I)e_t||_2^2 \quad (6)$$

- **Rescal** [31]: Following the Tensor Factorization methodology, the Rescal model transforms the Knowledge Graph's (head, relation, tail) triple into a three-way tensor, $X$. Here, $X$ is of dimensions (n*n*m), where n is the number of entities, m is the number of relations, and $X_{ijk} = 1$ indicates that there is a relation between entities. RESCAL model computes a score for a triple using the formula as in (7):

$$f_r(h, t) = h^T M_r t \quad (7)$$

where $h, t \in R^d$ are embeddings of head and tail, and $M_r \in R^{d \times d}$ is matrix form of relations.

- **RotatE** [32]: Before RotatE, it was not feasible to describe complicated relations like symmetry/antisymmetry, inversion, and composition in a Knowledge graph. The RotatE model follows Euler's identity $e^{i\vartheta} = \cos\vartheta + i\sin\vartheta$ and is designed to map entities and relations onto a complex vector space. It accomplishes this by defining each relation as a rotation from the source entity to the target entity. RotatE follows the score function as given in (8):

$$f_r(h, t) = -||h \circ r - t||^2 \quad (8)$$

where $\circ$ represents hadamard product.

2) Pointwise based Training KGE Models-

- **ComplEx** [33]: ComplEx embedding strategies attempt to address the problem of representing antisymmetric relationships in the Knowledge Graph. It accomplishes this by calculating the Hermitian dot product of head and tail entities. The score function it uses is given by (9):

$$f_r(h, t) = Re(\langle h, r, \bar{t} \rangle) \quad (9)$$

where $\langle . \rangle$ denotes the generalized dot product and $\bar{\cdot}$ denotes the conjugate of complex vectors.

- **DistMult** [34]: The DistMult embedding model is a simplified version of the Rescal model, which involves transforming the relation matrix, Mr, into a diagonal matrix. As a consequence, there is a decrease in the number of parameters, leading to increased scalability and improved performance. The score function is thus given in (10):

$$f_r(h, t) = h^T diag(r)t \quad (10)$$

- **SimplE** [35]: The SimplE embedding model improves upon the canonical polyadic decomposition method of tensor factorization by obtaining two interdependent embeddings for each entity. The scoring function of SimplE is defined in (11) as the average of triples $(h_i, r, t_j)$ and $(h_j, r^{-1}, t_i)$, i.e.

$$f_r(h, t) = \frac{1}{2}(\langle h_i, r, t_j \rangle + \langle h_j, r^{-1}, t_i \rangle) \quad (11)$$

where $r^{-1}$ represents inverse of relation r.

3) Projection-Based (Multiclass) Training KGE Models

- **TuckER** [36]: TuckER is a generalized version of the linear models Rescal, DistMult, ComplEx, and SimplE. TuckER uses the tucker decomposition of binary tensors to represent (h,r,t) triples in a knowledge graph. The scoring function followed by TuckER is given in (12):

$$f_r(h, t) = W \times_1 h \times_2 w_r \times_3 t \quad (12)$$

where $\times$ denotes tensor product along the $i^{th}$ mode and $W \in R^{n*n*m}$ (n-number of entities, m-number of relations).

The Algorithm 1 gives the working of our proposed algorithm. Pykg2vec makes the computational time shorter by generating mini-batches through the utilization of multi-processing units. We pass the *(Head, Tail, Relation)* KG file into the KG controller, which looks after its parsing tasks and generates the training, test, and validation sets. The control then moves onto the Batch Generator that makes queues of the mini-batches. These mini-batches are then processed by the core models that contain state-of-the-art Knowledge Graph embedding algorithms. Each of their modules has a loss function along with embedding operations.

The models are supplied with a configuration file that gives the necessary information to parse the datasets along with the baseline hyperparameters. The next part is the Training module which takes an instance of the KGE model and trains the model. The last part is the Evaluator that performs link prediction and provides the accuracy scores of the models. A Bayesian hyperparameter optimizer helps find the optimal golden hyperparameter by minimizing the loss function and improving the accuracy of our model. This Bayesian optimizer performs better than the brute force approaches and hence helps in significantly reducing the computational time.

The general principle involves learning the relations and entities that have been represented as facts. A negative sampling technique is followed where a chunk of negative triplets gets sampled from the set of positive triples by corrupting their entities. A scoring function is used to punish the negative triples and praise the positive triples. The maximizing and minimization of the scoring function is determined by the Bayesian optimization algorithm where the KG methods get evaluated based on their ability to predict the missing entity values in the negative triples *(?, r, t)* or *(h, r, ?)*. We perform 2 experiments for performing the task of Link prediction. In the $1^{st}$ experiment we train our Knowledge Graph across different embedding models given in Table 7. In experiment-2 we perform a set of 4 sub-experiments where we drop a group of similar attributes and analyze its impact on the task of Link Prediction as shown in Table 9 and Table 10.

### 2) SAppKG-D
The shallow embedding approaches suffer from drawbacks; they lack syntactic representation. To alleviate these limitations, shallow encoders can also be replaced with more sophisticated encoders that depend more generally on the structure and attributes of the graph. Some encoders can be generalized beyond the shallow embedding. For instance, the encoder can use node features or the local graph structure around each node as an input to generate an embedding. The key idea is that we want to generate representations of nodes that actually depend on the structure of the graph, as well as any feature information possessed by the graph. This structural information can be useful for many tasks.

In this paper, we propose a novel approach for app recommendation using knowledge graph embeddings. Our model, called SAppKG-D, combines a shallow embedding approach with a relation-weighted graph convolution-based deep embedding technique to extract higher-order semantic information from the knowledge graph. We address the limitations of shallow embeddings by using an encoder that depends on the graph structure and attributes to generate node embeddings that incorporate the structural information of the graph. Specifically, we use TransD embeddings and employ a neighbor aggregation process to propagate information between nodes in the graph, capturing higher-order connectivity and relations between entities. Our model is trained using a negative sampling strategy and the Adam optimizer to learn the recommendation model parameters. The model is evaluated using precision, recall, and mean average precision (MAP-N) metrics. The Algorithm 2 gives the working of our SAppKG-D model. SAppKG-D is a graph convolution-based deep embedding technique that is used to generate node embeddings. Mathematically, the model is represented in (13):

$$W_e = [w_{e,s}; w_{e,d}] \in \mathsf{R}^{ds+d_d} \qquad (13)$$

here, $w_{e,s}$ and $w_{e,d}$ represent the shallow and deep embeddings, respectively.

To incorporate the influence of neighboring apps, we employ an aggregation process in our proposed graph. In the knowledge graph (KG), the user-specific app attributes, such as installs, ratings, and other relevant user-centric information, dynamically change with time and are influenced by the availability of other similar competitor apps. Therefore, our model aggregates information from tail apps to head apps, taking into account the user-attributes associated with each app. Additionally, we introduce a weighted aggregation mechanism that considers the side information of the neighboring apps, enabling us to capture the relevant user-centric attributes and characteristics of neighboring apps. This approach enhances the overall representation and understanding of each app within the graph, considering the impact of user preferences and interactions Specifically, given an app $app1$ and a node $app2$ in the ARKG ($\mathcal{G}$), we define $N_{app2} = \{(h, r, t) | (h\ app2)\ (h, r, t) \quad \wedge \quad \in G\}$ as the set of triplets where $app2$ is the head entity. The aggregated vector of neighbors for $app2$, specific to $app1$, is computed as follows in (14):

$$\mathbf{v}_{app1}^{app2} = \sum_{(h,r,t) \in N_{app2}} w_{app1}^r \mathbf{t} \qquad (14)$$

here, $\mathbf{t} \in \mathsf{R}^d$ represents the vector of the tail entity $t$, and $w_{app1}^r$ is the weight between app $app1$ and relation $r$, characterizing the importance of relation $r$ to app $app1$. The weight $w_{app1}^r$ is computed as follows in (15):

$$w_{app1}^r = \frac{\exp(\pi(\mathbf{app1}, \mathbf{r}))}{\sum_{(h,r,t) \in N_{app2}} \exp(\pi(\mathbf{app1}, \mathbf{r}))} \qquad (15)$$

In the above equations, $\mathbf{app1} \in \mathsf{R}^d$ and $\mathbf{r} \in \mathsf{R}^d$ represent the embeddings of app $app1$ and relation $r$, respectively. The function $\pi : \mathsf{R}^d \times \mathsf{R}^d \rightarrow \mathsf{R}^d$ refers to a weight score function that maps two vectors from the Euclidean space $\mathsf{R}^d$ to a single

---

**Algorithm 2 SAppKG-D**

**Require:**
 $G = E, R$ : knowledge graph
 $E_{app}$ : app embedding matrix
 $D_{train}$ : training data set
 $D_{test}$ : test data set
 $\lambda_1, \lambda_2$ : hyperparameters
 $d$ : hidden dimension
**Ensure:**
 $W_e$ : entity embedding matrix
 $W_r$ : relation embedding matrix
1: $E_{train} \leftarrow \text{ExtractSamples}(D_{train}, W_e, W_r, P)$            ▷ sample extraction
2: $E_{test} \leftarrow \text{ExtractSamples}(D_{test}, W_e, W_r, P)$            ▷ sample extraction
3: $loss \leftarrow \text{TransD}(E_{train}, \lambda_1, \lambda_2)$         ▷ embedding training with TransD
4: $W_e, W_r \leftarrow \text{TransD}(G, d)$          ▷ graph embedding propagation
5: $P \leftarrow \text{Concatenate}(P_{app}, P_{pkg})$         ▷ embedding concatenation
6: $precision, recall, map_n \leftarrow \text{Metrics}(E_{test}, W_e)$         ▷ evaluation metrics

---

vector in $\mathbb{R}^d$. In the context of the research paper, this function is used to calculate the weights between the app $app1$ and the relation $r$.

To seamlessly update the embeddings for the next layer, we perform the following steps. For each app node $app1$, we concatenate its current representation $\mathbf{v}_{app1}$ with the aggregated vector of its neighboring apps $\mathbf{v}_{app1}^{N_{app2}}$. This concatenation captures the combined information from the app itself and its neighboring apps. This concatenated vector is then passed through a fully connected layer with a nonlinear activation function $\sigma$, which transforms it into the new representation of $app1$. The update process can be formulated as follows as given in (16):

$$\mathbf{v}'_{app1} = \sigma \left( \mathbf{W} \cdot \left( \mathbf{v}_{app1} \| \mathbf{v}_{app1}^{N_{app2}} \right) + \mathbf{b} \right) \quad (16)$$

here, $\mathbf{v}'_{app1}$ (i.e., the output of this layer) represents the new representation of the app node $app1$ specific to its connections with other apps. The transformation weight and bias are denoted as $\mathbf{W}$ and $\mathbf{b}$, respectively. The symbol ''$\|$'' denotes the concatenation operation, which combines the current representation of $app1$ with the aggregated vector of its neighboring apps. This process ensures that the embeddings are continuously updated and refined as the information flows through the layers.

After updating the embeddings for each app node in the previous step, our next objective is to incorporate higher-order connectivity, which plays a crucial role in improving the quality of recommendations. To achieve this, we leverage the concept of information propagation among different layers of the knowledge graph. This enables us to capture higher-order structural proximity among entities and enhance the representation of their relationships.

In our approach, we stack multiple propagation layers, typically $K$-1 layers, when we are at the $K^{th}$ level. This process allows us to propagate information and update the embeddings based on aggregated information from neighboring entities. By iteratively aggregating and incorporating information from multiple layers, we can capture and integrate higher-order dependencies and structural patterns within the knowledge graph, leading to more accurate and comprehensive recommendations.

To formalize this process, we utilize Equation (16) for propagating embeddings along higher-order connectivity. For convenience, we denote the representation of node $v$ specific to $app1$ at depth $k$ - 1 as $\mathbf{v}_{app1}^{(k)}$, which combines the initial representations of node $v$ and its neighbors up to $k$ hops away. This information propagation technique is illustrated in the SAppKG-D: Training and Evaluation subsection having the information propagation image in Figure-3. By incorporating higher-order connectivity, we can effectively capture the structural proximity between entities and enhance the recommendation process. Through a series of information propagation layers, our model updates the representations of app nodes to capture their higher-order dependencies and long-range inter-relatedness. The final embedding $\mathbf{v}_{app1}^{(K)}$ of an app node $v$ specific to app $appI$ reflects its structural connections up to $K$ hops. Additionally, leveraging the general knowledge graph (KG) embeddings, we gain insights into the relational distances between entities.

To predict interactions between apps $appI$ and $appII$, we combine their embeddings into a unified vector as shown in Equation (17):

$$\mathbf{a}^*_{app1} = \mathbf{a}_I \| \mathbf{a}_{app1}^{(K)} \quad (17)$$

In this equation, the vectors $\mathbf{a}_I$ capture the relationship-specific information and the characteristics of app $appI$ derived from the KG embeddings. On the other hand, the embedding $\mathbf{a}_{app1}^{(K)}$ represents the final output specific to app $appI$, obtained through the convolutional embedding propagation component. By concatenating these vectors, we create a unified representation that combines both relationship-specific information and the learned characteristics of the app, enabling us to predict their interaction. The matching score between app $appI$ and app $appII$ is computed by taking the inner product of their embeddings as shown in (18):

$$y_{app1, \, app2} = \mathbf{a}^*_{app1} \cdot \mathbf{a}^*_{app2} \quad (18)$$

To train our app recommendation model, we employ negative sampling and utilize a binary cross-entropy loss function with

$L_2$ norm regularization as shown in (19):

$$L_{CEP} = \sum_{app1 \in A} \sum_{app2 \in Trn^{app1}} \left[ -\log(y_{app1, app2}) \right.$$
$$+ \sum_{i \in Neg^{app1}_{app2}} \left. -\log(1 - y_{app1, i}) \right] + \lambda \|\vartheta\|^2_2 \quad (19)$$

here, $A$ represents the set of all apps. $Trn^{app1}$ denotes the training instances involving app *app1*, and $Neg^{app1}app2$ represents the randomly sampled negative app instances associated with app *app1* and app *app2*. The term $y_{app1, app2}$ denotes the predicted probability of a positive interaction between app *app1* and app *app2*, while $y_{app1, i}$ represents the predicted probability of a positive interaction between app *app1* and a negative app instance $i$. The first part of the loss function corresponds to the log-likelihood of positive interactions, while the second part captures the log-likelihood of negative interactions. In addition, the regularization term $\lambda|\vartheta|^2_2$ is included to prevent overfitting. Here, $\lambda$ represents the regularization coefficient, and $\vartheta$ denotes the model parameters.

To optimize the model, we use the Adam optimizer, which is a popular optimization algorithm known for its efficiency and effectiveness in training deep neural networks. By minimizing the loss function (19) using negative sampling and regularization, our model learns to make accurate predictions and generate meaningful app recommendations

To assess the effectiveness of our model, we utilize several performance metrics, namely precision (P), recall (R), and mean average precision (MAP-$N$). Precision is calculated as the ratio of the number of recommended relevant apps (TP) to the total number of recommended apps (TP + FP), and can be expressed as (20):

$$P = \frac{TP}{TP + FP} \quad (20)$$

Recall measures the ratio of the number of recommended relevant apps (TP) to the total number of relevant apps (TP + FN), and is given by equation (21):

$$R = \frac{TP}{TP + FN} \quad (21)$$

Mean average precision (MAP-$N$) considers the precision at each relevant position in the top-$N$ recommended apps and calculates the average across all relevant apps. It can be formulated as (22):

$$MAP\text{-}N = \frac{1}{|R|} \sum_{app \in A_{rec}} \frac{Prec(app)}{Rank(app)} \quad (22)$$

In this equation, $R$ represents the set of relevant apps, $|R|$ denotes the total number of relevant apps, and $A_{rec}$ represents the set of recommended apps. Prec(app) denotes the precision at each relevant app position, and Rank(app) represents the relevance position of each relevant app in the recommended list. The equation captures the essence of MAP-$N$, which evaluates the average precision of relevant apps considering their positions in the recommended list. These metrics provide a comprehensive evaluation of the model's performance in generating relevant app recommendations, considering both the accuracy and comprehensiveness of the recommendations.

## IV. EXPERIMENTAL SETUP

We scraped the data from Google Play Store for different categories to train the model over a diverse set of data categories as reported in Table 6. *Photography* and *Productivity*, contain the extracted apps from four subcategories namely *Top Free*, *Top Paid*, *Grossing* and *Trending*. And for the *Games*, the extracted apps are from five sub-categories namely *Top Free*, *Top Paid*, *Grossing*, *New Free*, and *New Paid*. The *New Free* and *New Paid* contain apps that have been launched within a month from today's date. As the Games genre didn't have a Trending category we extracted the New Free and New Paid categories.

**TABLE 6.** Scraped apps.

| | Photography | Productivity | Games |
|---|---|---|---|
| TOP FREE | 200 | 200 | 200 |
| TOP PAID | 104 | 151 | 200 |
| GROSSING | 200 | 200 | 200 |
| TRENDING | 200 | 200 | - |
| NEW FREE | - | - | 200 |
| NEW PAID | - | - | 3 |
| TOTAL | 704 | 751 | 803 |
| TOTAL(After removing duplicates) | 509 | 509 | 775 |

The metrics used for evaluating the Link Prediction task include the following:

1) *Mean Rank [37]*- Lists out the rank of the answer in the predicted list and then displays their mean result as shown in (23).

$$MR = \frac{1}{|Q|} \sum_{i=1}^{|Q|} rank_i \quad (23)$$

where Q is a set of triples and $rank_i$ denotes the rank of a such triple.

2) *Mean Reciprocal Rank [37]*- Lists out the mean of the inverse of all the ranks as in (24)

$$MRR = \frac{1}{|Q|} \sum_{i=1}^{|Q|} \frac{1}{rank_i}, \quad (24)$$

3) *Hits@K [37]*- Ratio of answers ranked in the list of top-k elements as shown in (25)

$$Hits@K = \frac{|\{i \in Q \mid rank_i \leq K\}|}{|Q|} \quad (25)$$

4) *Filtered Hits@K*- Removing the corrupted triplets included in the train, valid, and test sets before ranking is known as *Filter*. The ratio of answers ranked in top-k after performing this filter is known as *Hits@K*. The formula of computation remains the same as in (25)

## V. EXPERIMENTAL RESULTS

### A. SAppKG-S

SAppKG-S is a shallow embedding-based methodology that proposes apps using shallow embedding models in a knowledge graph while also providing comparisons between various such models through the metrics of MR, MRR, and Hits@K.

#### 1) EXP-1

The Table-7 shows the Mean Ranking, Mean Reciprocal Ranking, and Hits@K scores for the different embedding models. A low score of Mean Ranking is assumed to be better, and from Table-7, we can observe that the RotatE model has the lowest mean ranking score and hence the best performance followed by Translation based models TransH and TransD.

The column of mean reciprocal ranking has a score range of 0 to 1. It is observed that the model of Complex has a score of 0.668 which is the nearest to 1 hence the best among all the other models. This is followed by RotatE and CP. The next 4 columns have a score of Hits@(1,3,5,10). It can be observed RotatE and ComplEx perform the best. These models have scores in the range of 0.6 - 0.8 followed by TransH. DistMult, CP, KG2E, and TransD. We compare the Model performance by taking their averages across all the hits.

From the Table 7, we can conclude that the neural network. and projection-based models perform the worst as they are computationally costly. RotatE performs the best as it replaces the traditional translation-based operations with a rotation based operation helping to distinguish different relations like antisymmetry. composition, and symmetry. RotatE outperforms the other models indicating that the rotation-based operation helps model the different relation types and their semantic aspects [38].

Compared to translational-based models TransE, TransD, TransH DistMult, and other models. The Complex, RotatE advantage through multi-part high-dimensional embeddings to achieve state-of-the-art performances on link prediction. All parts of the embeddings get adjusted simultaneously in these models [39]. Model Optimization and selection of the appropriate loss while training stands of utmost importance: most distance-based and translation-based models like TransE, TransD and TransH use a margin rank loss function. ComplEx uses a binary logistic loss function to obtain the best results. RotatE tweaks the loss function without the regularization term by adding a margin parameter that significantly enhances the performance [39].

#### 2) EXP-2

After obtaining the model training scores and performances, we evaluate the effect of the trained KG attributes on the KG performance. For conducting this experiment, we group the attributes into 4 group of categories labelled as ex 1, ex 2, ex 3 and ex 4. Attributes that convey a similar category or type of side information and are highly correlated with one another are grouped together. The Table 8 given below shows the list of attributes that have been grouped together along with their grouping description. The grouped attributes are dropped from the Knowledge Graph and the modified Graph is trained on the rest of the attributes and its performance is noted for Rotate and ComplEx embeddings. We train all the 4 modified graphs on RotatE and ComplEx, since the performance of the original graph is the best for these 2 models. Hence, we limit our training analysis to these 2 models. Table 9 given below shows the performance of the 4 modified graphs on ComplEx, We can find that dropping released and size attributes in Exp 3 improves the performance of the Complex model probably as these attributes might be causing noise in the predictions. Dropping the features in Exp 4 doesn't change much from the original model performance and the model performance remains the same. Dropping the features in Exp 2 leads to a slight decrease in the performance of the overall model. Dropping the attributes in Exp 1 significantly decreased our model performance making it close to zero. Hence, we can conclude that dropping the *released, size* attributes and preserving the binary features would improve our model performance on the Complex model.

The presented Table 10 highlights the performance of the modified graphs on RotatE. Based on the table, we can conclude that dropping the attributes in Exp 4 leads to an improvement in the model's performance. On the other hand, dropping the features in Exp 3 does not significantly alter the performance of the original model performance. Additionally, dropping the features in Exp 2 leads to a slight decrease in the overall model performance, whereas dropping the attributes in Exp 1 causes a moderate decrease in the model's performance. Therefore, it can be concluded that preserving the binary feature group would improve the model performance, and this result aligns with the ComplEx model results. In summary, the results suggest that certain attribute groups can be dropped to improve the model's performance, while preserving other groups, such as the binary features, can positively impact the model's performance.

#### 3) EXP-3

In our third experiment, we evaluated the performance of the User-App Knowledge Graph dataset, used by [5], on the SAppKG-S model. Our results show that the mean ranking and mean reciprocal ranking of our Knowledge graph outperform the User-App KG for the TransD model. Although the User-App KG performs better than our KG for Hits@1, our KG stands out from the User-App KG for Hits@(3,5,10). The comparison results of SAppKG-S and KGEP are shown in Table 11. Specifically, the filtered mean rank and MRR of our SAppKG-S model are better than those of the KGEP model. Additionally, our SAppKG-S model outperforms the KGEP model for Hits@(3,5,10), while the KGEP model performs slightly better than our SAppKG-S model for Hits@1.

### B. SAppKG-D

SAppKG-D combines a shallow embedding approach and a relation-weighted graph convolution-based deep embedding

**TABLE 7.** Training Results.

| Model | Filtered MR | MRR, Filtered MRR | Hits1 | Hits3 | Hits5 | Hits10 |
|---|---|---|---|---|---|---|
| Complex | 107.6109 | **0.6684** | **0.5918** | **0.737** | 0.7574 | 0.7767 |
| CP | 334.0777 | 0.4537 | 0.3725 | 0.5215 | 0.5495 | 0.5759 |
| DistMult | 143.8794 | 0.355 | 0.1639 | 0.5212 | 0.5876 | 0.6403 |
| KG2E | 159.3505 | 0.1962 | 0 | 0.31 | 0.472 | 0.6074 |
| NTN | 309.7834 | 0.0926 | 0.0374 | 0.0918 | 0.13 | 0.2042 |
| Rescal | 121.0371 | 0.0894 | 0.0286 | 0.0808 | 0.1248 | 0.2095 |
| RotateE | **34.7446** | 0.5053 | 0.3172 | 0.6391 | **0.7644** | **0.8637** |
| SimplE | 895.9388 | 0.0049 | 0.0007 | 0.002 | 0.0037 | 0.0069 |
| TransD | 88.5658 | 0.1995 | 0.0008 | 0.2876 | 0.475 | 0.6471 |
| TransE | 125.9044 | 0.2101 | 0 | 0.3442 | 0.5062 | 0.6324 |
| TransH | 73.1269 | 0.2341 | 0 | 0.3635 | 0.5725 | 0.7509 |
| TuckER | 913.3816 | 0.0036 | 0.0001 | 0.0006 | 0.0016 | 0.0047 |

**TABLE 8.** Feature Grouping.

| Relations | Grouping Description | Experiment no. |
|---|---|---|
| adSupported, editorsChoice, offerIAP, video | Binary features are grouped into one category | Exp 1 |
| contentRating, genreId | Multiple features related to its content and genre are grouped into one category | Exp 2 |
| released, size | Features related to other attributes like size and release date are grouped together | Exp 3 |
| reviews, installs, scoreText, ratings | Features related to app interactions on the google play store are grouped into one category | Exp 4 |

**TABLE 9.** ComplexE results.

| Complex | Filtered MR | MRR, Filtered MRR | Hits1 | Hits3 | Hits5 | Hits10 |
|---|---|---|---|---|---|---|
| Exp 1 | 874.967 | 0.0034 | 0 | 0 | 0.001 | 0.003 |
| Exp 2 | 234.278 | 0.4973 | 0.4304 | 0.5435 | 0.5707 | 0.6009 |
| Exp 3 | 65.8182 | 0.6896 | 0.5948 | 0.7641 | 0.8058 | 0.8477 |
| Exp 4 | 241.791 | 0.6637 | 0.6206 | 0.7045 | 0.7111 | 0.7178 |

**TABLE 10.** RotateE Results.

| RotateE | Filtered MR | MRR, Filtered MRR | Hits1 | Hits3 | Hits5 | Hits10 |
|---|---|---|---|---|---|---|
| Exp 1 | 23.744 | 0.3289 | 0.15 | 0.403 | 0.5475 | 0.725 |
| Exp 2 | 16.7458 | 0.3332 | 0.127 | 0.444 | 0.5948 | 0.7637 |
| Exp 3 | 12.2357 | 0.4485 | 0.2338 | 0.5913 | 0.7313 | 0.8749 |
| Exp 4 | 37.7873 | 0.688 | 0.5844 | 0.7555 | 0.8179 | 0.8843 |

technique to extract higher-order semantic information from the knowledge graph to propose apps.

### 1) EXP-4

In our fourth experiment, we evaluated the performance of our Knowledge Graph (KG) on the Graph Convolutional Network (GCN) used by Zhang et al. [5] for training and testing the User-App dataset. To establish a baseline for comparison, we used their results as our reference. Table 12 shows that our KG performed slightly worse in terms of precision, with a margin of—8% to–20% compared to the standard baseline, we saw a positive deviation margin ranging from 15% to 49% in the recall. In terms of the Mean Average Precision (mAP-N) of our model performed slightly better, with a range of up to 7%.

To achieve the best possible results, we experimented with various hyperparameters and identified the following settings as optimal: aggregator type as concat, a neighbour sample size of 7, embedding dimensions of 16, one iteration for computing entity representation, a batch size of 10, L2 regularizer weight of 1e-7, and a learning rate of 0.005. We trained our model for 200 epochs, using a training/test/validation split ratio of 0.6, 0.2, and 0.2, and achieved the best result at the 69th epoch.

To evaluate the performance of our graph, we perform relation prediction using SAppKG-D. As we only have 11 relations, we make predictions for 1, 3, 5, and 7 relations and measure precision, recall, and MAP-N for each. We find that all three metrics improve linearly as the number of predicted relations increases, with the best results obtained for predicting 7 relations.

For relation prediction, we use the hyperparameters: aggregator type of concat, embedding dimension of 16, one iteration for computing entity representation, a batch size of 10, L2 regularizer weight of 1e-7, the learning rate of 0.005, and neighbour sample size of 5. We use a training/test/validation split ratio of 0.6/0.2/0.2 and train for 200 epochs. Table 14 shows that with these hyperparameters we achieve a precision of 0.291, recall of 0.622, and MAP-N of 0.383, representing a significant improvement over our baseline results.

### 2) EXP-5

In our fifth experiment, we focused on evaluating the average inference time of the SAppKG-S and SAppKG-D models for entity prediction on the Google Play Store and Apple App Store datasets. To ensure robustness and facilitate a fair comparison between Android and iOS apps, we adhered to a standardized data scraping and knowledge graph construction process, as described in the methodology section. This approach allowed us to curate a dataset comprising 1793 apps, carefully selected to match the size and categories of the Google Play Store dataset. To assess the performance of our model, we conducted experiments using a test set consisting of 20% of the total apps in our dataset, resulting in 358 apps. In order to measure the average inference time for each model (SAppKG-S with ComplEx, SAppKG-S with RotatE, and SAppKG-D), we employed Equation 26.

**TABLE 11.** Comparison of SAppKG-S and KGEP.

| Model | Filtered MR | MRR | Hits1 | Hits3 | Hits5 | Hits10 |
|---|---|---|---|---|---|---|
| TransD (SAppKG-S) | **88.5658** | **0.1995** | 0.0008 | **0.2876** | **0.475** | **0.6471** |
| TransD (KGEP) | 2428.304 | 0.0270 | **0.0105** | 0.0235 | 0.0333 | 0.0524 |

**TABLE 12.** Comparison of SAppKG-D and KGEP.

| Number of app predicted(K) | Precision | | | Recall | | | Map-N | | |
|---|---|---|---|---|---|---|---|---|---|
| | SAppKG-D | KGEP | Dev(%) | SAppKG-D | KGEP | Dev(%) | SAppKG-D | KGEP | Dev(%) |
| 10 | 0.8 | **1** | -20 | **2.833** | 2.46 | 15.13 | **4.093** | 3.853 | 5.86 |
| 20 | 0.55 | **0.6** | -8.3 | **4.583** | 3.061 | 49.73 | **4.306** | 3.996 | 7.21 |
| 30 | 0.433 | **0.567** | -23.57 | **5.416** | 4.256 | 27.27 | **4.388** | 4.177 | 4.82 |
| 40 | 0.35 | **0.475** | -26.32 | **5.75** | 4.839 | 18.83 | **4.418** | 4.232 | 4.22 |

**TABLE 13.** Relation prediction SAppKG-D.

| Number of relations predicted (K) | Precision | Recall | Map-N |
|---|---|---|---|
| 1 | 0.170 | 0.110 | 0.170 |
| 3 | 0.187 | 0.213 | 0.281 |
| 5 | 0.228 | 0.442 | 0.357 |
| 7 | **0.291** | **0.622** | **0.383** |

$$\bar{T} = \frac{1}{N} \sum_{i=1}^{N} T_i \quad (26)$$

here, $\bar{T}$ represents the average inference time, $N$ denotes the total number of apps in the test set and $T_i$ represents the total time taken to predict each app.

We specifically chose the ComplEx and RotatE models for these experiments as they exhibited superior performance in our previous SAppKG-S experiments. Additionally, we included the SAppKG-D model for comparative analysis. Table 14 displays the average inference times of various models, namely SAppKG-S with ComplEx embeddings, SAppKG-S with RotatE embeddings, and SAppKG-D, for both the Play Store and the App Store. For the SAppKG-S model using ComplEx embeddings, the average inference time is recorded as 18.149ms for the Play Store and 16.625ms for the App Store. This represents a percentage difference of −7.55% (Play Store) and − 4.09% (App Store) compared to the SAppKG-S model with RotatE embeddings, which has average inference times of 15.413ms (Play Store) and 15.147ms (App Store).

The SAppKG-S model, on average, has an inference time of 16.781ms for the Play Store and 15.886ms for the App Store. Comparing this to the SAppKG-D model, which has average inference times of 27.077ms (Play Store) and 29.487ms (App Store), we observe a percentage difference of 61.02% (Play Store) and 85.63% (App Store) respectively. These percentage differences highlight the varying computational efficiency among the models, with negative differences indicating faster inference time.

**TABLE 14.** Average Inference time: SAppKG-S.

| Model | Play Store | App Store |
|---|---|---|
| SAPPKG-S: ComplEx | 18.149ms | 16.625ms |
| SAPPKG-S: RotatE | 15.413ms | 15.147ms |
| SAPPKG-S: Average Inference Time | 16.781ms | 15.886ms |
| SAPPKG-D | 27.077ms | 29.487ms |

## VI. CONCLUSION

An efficient mobile app recommender system has the potential to save users time and effort in selecting the best app, while also providing app developers and stakeholders with the benefits of increased revenue and profits from more app downloads. Overall, our research contributes to the field of mobile app recommendation by introducing a novel knowledge graph-based approach that effectively incorporates side information and maintains user privacy while achieving improved performance compared to baseline models. In this paper, we developed a Knowledge Graph (KG), starting with data scraping, analysis, and constructing the Head, Tail, and Relationship tables. We described the methodology for preprocessing and mapping each relation and analyzed the statistical characteristics of the KG. We also built different embeddings using Knowledge Graph Embedding methods and proposed a novel approach called SAppKG-S that utilizes shallow embedding models in a knowledge graph to recommend apps while providing comparisons between different models. To evaluate the performance of our approach, we used the embeddings to perform the task of link prediction and compared their performance to various state-of-the-art models. Our KG has twelve relations, and we found that modelling these many relations would be a tedious task. However, we observed that RotatE and ComplEx performed the best in this task due to their high-dimensional multi-art embeddings used to model the different relations.

We also proposed a hybrid approach called SAppKG-D, which combines shallow embedding with relation-weighted graph convolution-based deep embedding techniques to extract higher-order semantic information from the KG for app recommendations. Our results showed that our approach outperformed the baseline results. Furthermore, we found that binary attributes play an important role in preserving the structural information of the KG, and removing user data from the KG while incorporating app attributes in the form of edges gives similar results to the state-of-the-art KGEP model while maintaining user privacy. Overall, our work demonstrates the effectiveness of Knowledge Graph Embedding methods in building a mobile app recommender system and highlights the potential of our proposed approaches, SAppKG-S and SAppKG-D, for improving the performance of app recommendations.

In the future, we plan to enhance our model by increasing the graph inter-connectedness to further increase density. Given that we were limited by computational power, the graph was trained for only 500 epochs, so we can also explore training it for longer to see if performance can be improved. We also plan to increase the number of relationships and entities in the KG and connect nodes that match entities to make the graph denser and less sparse. These improvements have the potential to further enhance the performance and utility of our proposed model.

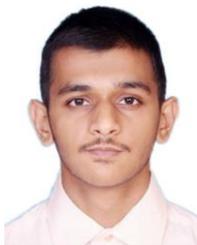

**DAKSH DAVE** received the B.E. degree (Hons.) in electrical electronics from the Birla Institute of Technology and Science, Pilani (BITS Pilani), Pilani Campus, Rajasthan, India, in 2022. From 2019 to 2022, he was a Research Assistant with the Internet of Things Laboratory, Department of Humanities, and the Department of Computer Science, BITS Pilani. He was also a Data Scientist with Samsung Research and Development, L&T Chiyoda, and SimplerCRM. Later, he was a Backend Developer with Eltropy and Expound Technivo. His research interests include recommender systems, knowledge graphs, data mining, and 5G and beyond technologies.

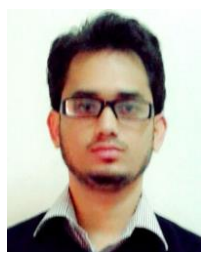

**ADITYA SHARMA** received the B.Tech. degree in information technology engineering from Baba Ghulam Shah Badshah University, Jammu and Kashmir, in 2014, and the M.Tech. degree in computer science engineering from Shri Mata Vaishno Devi University, Katra, Jammu and Kashmir. He is currently pursuing the Ph.D. degree in computer science engineering from the Birla Institute of Technology and Science, Pilani, Rajasthan, India. He is with the Department of Education, Government of Jammu and Kashmir. His current research interests include graph neural networks and their applications in natural language processing.

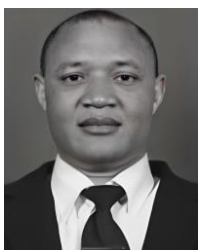

**SHAFI'I MUHAMMAD ABDULHAMID** is currently an Assistant Professor with the Department of Information Technology, Community College of Qatar. His research interests include soft computing, machine learning, fog computing, and cloud computing security.

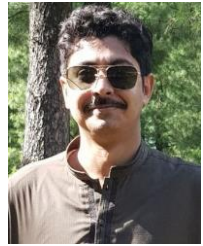

**ADEEL AHMED** received the Ph.D. degree in computer science from Quaid-i-Azam University, Islamabad, Pakistan, in 2022. He was with the software industry for some years. He is currently a Faculty Member with the Department of Information Technology, The University of Haripur, Khyber Pakhtunkhwa, Pakistan. His research interests include natural language processing, machine learning, recommendation systems, social network analysis, and information visualization.

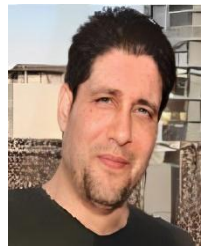

**ADNAN AKHUNZADA** (Senior Member, IEEE) is currently a Cybersecurity Specialist and a Consultant with extensive industrial experience and knowledge. He has over a decade of professional research and development experience with a proven track record of high-impact published research and innovations, such as U.S. patents, design, and development of commercial cybersecurity products and frameworks. He has advised some of the largest companies in the world, assuring security on multi-million projects. He has served in several industrial and academic positions with leading players in cyber security. His expertise lies in designing novel SIEM systems, IDS, IPS, threat intelligence platforms, secure protocols, AI for cybersecurity, secure future internet, adversarial and privacy preserving machine learning, and QoS/QoE of emerging computational and communication networks.

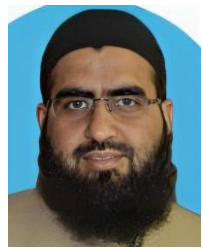

**RASHID AMIN** received the M.S.C.S. and M.C.S. degrees from International Islamic University, Islamabad, and the Ph.D. degree in computer science from COMSATS University Islamabad, Wah Campus, Pakistan. He is currently an Assistant Professor with the Department of Computer Science, University of Chakwal, Pakistan. Before this, he was a Lecturer with the Department of Computer Science, University of Engineering and Technology, Taxila, Pakistan, for seven years, and the University of Wah, Wah Cantt, Pakistan, for four years. He supervised many M.S. degree level student's thesis, and five Ph.D. degree students are under his supervision. He has published several research papers on ML, DL, and SDN in well-reputed venues, like IEEE Communications Surveys and Tutorials, IEEE Access, *Electronics* (MDPI), and *IJACSA*. His current research interests include machine learning, deep learning, the IoMT, distributed systems, and cyber security. He is co-editing some Special Issues in some renowned journals. He is a Reviewer of international journals, such as NetSoft, LCN, IEEE Globecom, FiT, IEEE Wireless Communication, IEEE Internet of Things Journal, IEEE Journal on Selected Areas in Communications, IEEE Access, and IEEE System Journal.

• • •